\documentclass[a4paper]{panl}
\usepackage{cite}
\usepackage{wrapfig}

\usepackage{longtable}
\usepackage{rotating}
\usepackage{lscape}
\usepackage{multirow}
\usepackage{tabularx}

\usepackage{latexsym,graphicx,epsfig,psfrag,fancybox,xspace}
\usepackage{amssymb,amsmath,amsxtra,amsfonts}

\newcommand{\nn}{\nonumber\\}

\newcommand{\bea}{\begin{eqnarray}}
\newcommand{\ena}{\end{eqnarray}}
\newcommand{\be}{\begin{equation}}
\newcommand{\en}{\end{equation}}

\newcommand{\Br}{\ensuremath{\mathcal{B}}\xspace}

\newcommand{\sll}{/\kern-4pt l}
\newcommand{\slp}{p\kern-5pt/}
\newcommand{\sls}{s\kern-5pt/}
\newcommand{\slell}{/\kern-5pt\ell}

\originalTeX
\begin{document}
% Journal sections (see http://pkp.jinr.ru/index.php/PEPAN_LETTERS/about/editorialPolicies#focusAndScope)
%\issuearea{Physics of Elementary Particles and Atomic Nuclei. Theory}

\title{Anomalies in Weak Decays of Hadrons Containing a b Quark\\ }

\maketitle
\authors{Aidos Issadykov\vspace{1mm} $^{a,b}$ \vspace{2mm}\footnote{E-mail: issadykov@jinr.ru}}
\authors{Mikhail A. Ivanov\vspace{1mm} $^{a}$ \vspace{2mm}\footnote{E-mail: ivanovm@theor.jinr.ru}}

\from{$^{a}$\,Joint Institute for Nuclear Research, 141980 Dubna, \textit{RUSSIA}}
\from{$^{b}$\,The Institute of Nuclear Physics, \\
Ministry of Energy of the Republic of Kazakhstan, 050032 Almaty,  {\it KAZAKHSTAN}}

\begin{abstract}
A brief review of the current state of observed deviations of theoretical predictions from experimental
data in semileptonic decays of  $B$ and $B_c$ mesons is given. A theoretical analysis of these decays is carried
out, taking into account the effects of new physics, which appear due to the introduction of new four-fermion
operators, which are absent in the basis of the Standard Model (SM) operators. The necessary form factors
are calculated within the framework of the covariant quark model developed in our papers.
\end{abstract}
\vspace*{6pt}

\label{sec:intro}
\section*{INTRODUCTION}

To date, the Standard Model (SM) has been tested
and validated by many experiments. Attention has
shifted beyond the SM in search of new particles and
new interactions. It should be noted that so far it has
not been possible to observe new particles in modern
accelerators, in particular, at the Large Hadron Collider
at CERN. However, there are indirect indications
to the New Physics (NP) in a number of experiments
on the study of weak decays of hadrons containing
a b quark. Among them, semitau decays of b
hadrons should be noted ($B$,$B_s$, and $B_c$
 mesons and $\Lambda_b$ baryons), which are due to the transition $b\to c\tau\nu_\tau$, as
well as rare decays due to the transition $b\to s\ell^+\ell^-$.
The observed deviations of the experimental data in a
number of observed quantities from their values
obtained as part of the Standard Model have been
called “anomalies.” A common point of view among
theorists is that all these deviations can be explained by
the violation of lepton universality \cite{Crivellin:2022qcj,DiCanto:2022icc}. In other
words, the possible manifestations of the New Physics
lead to a difference in the interactions of muons, electrons,
and tau leptons with gauge bosons. This unified
point of view suggests a common origin of anomalies
from the point of view of physics beyond the SM,
which confirms the arguments in favor of the violation
of lepton universality. In particular, this path opens up
new possibilities for constructing NP models and
makes it possible to create a convincing physical justification
for experiments.

\label{sec:formfactors}
\section*{ANOMALIES AND VIOLATION
OF LEPTON UNIVERSALITY}
There are three generations of leptons in the SM,
grouped into lepton–neutrino doublets:
\bea
&&
\left( \begin{array}{c}
\nu_e \\[2ex]  e^- 
\end{array} \right), \qquad
\left( \begin{array}{c}
\nu_\mu \\ [2ex] \mu^- 
\end{array} \right), \qquad
\left( \begin{array}{c}
\nu_\tau \\[2ex]  \tau^- 
\end{array} \right).
\ena
The lepton universality in the SM means that, in
the interaction Lagrangian, the constants characterizing
the coupling of weak lepton currents to W bosons
are the same for all three generations of leptons. The
violation of lepton universality implies a difference in
the coupling constants for electrons, muons, and tau
leptons.\\

\noindent { \bf Semitauonic decays \boldmath{$b\!\!\to\!\! c\tau\nu$:}}

These decays are due to charged currents at the tree level in the SM and,
therefore, the decay branchings take values of the
order ${\mathcal O}(10^{-3})$. Differential decay width, ${\rm d}\Gamma$, for semileptonic decays involving $D^{(*)}$ mesons in the final state depends on both $m^2_{\ell}$ and $q^2$, of the square of the invariant mass of the lepton pair.
\begin{align}
  &\frac{{\rm d}\Gamma^{\rm SM}(\bar B\to D^{(\ast)}\ell^-\bar\nu_\ell)}
        {{\rm d}q^2}\,   
        =  \underbrace{\frac{G_F^2\; |V_{cb}|^2\;
            |\boldsymbol{p}^*_{D^{(*)}}| \; q^2}{96\pi^3 m_B^2}
  \left(1-\frac{m_\ell^2}{q^2} \right)^2}_{\text{universal and phase space  factors}}
 \nn & 
\times 
\underbrace{\left[(|H_{+}|^2+|H_{-}|^2
 +|H_{0}|^2) \left(1+\frac{m_\ell^2}{2q^2}\right) 
 + \frac{3 m_\ell^2}{2q^2}|H_{s}|^2 \right]}_{\text{hadronic effects}}~.
%\nonumber
\end{align}
Here, $\boldsymbol{p}^*_{D^{(*)}}$ is the momentum of the daughter hadron $B$ in the rest frame of the meson. Four helical amplitudes $H_\pm,H_0,H_s$ characterize the effect of the hadronic structure and depend on the square of the momentum transfer $q^2$, changing in the interval $m^2_{\ell} \le q^2 \le (m_B - m_{D^{*}})^2$.
The most optimal are measurements of the ratios of decay branchings with a tau lepton to branching with an electron (or muon) in the final state. In this regard, the dependence on the element of the Cabibbo–Kobayashi–Maskawa (CKM) matrix is eliminated $|V_{cb}|$, there is a partial reduction in theoretical uncertainties associated with hadronic effects, and experimental
uncertainties are reduced.
  \be
  R( D^{ (\ast)} ) =
  \frac
    {\Br( B \to D^{(\ast)}\tau \nu_\tau ) }
    {\Br( B \to D^{(\ast)}\ell \nu_\ell ) },
    \quad
    D^{ (\ast)} = D\,\,\text{or}\,\,D^\ast, \quad \ell = e\,\,\text{or}\,\,\mu
  \en
The predictions obtained in the SM are as follows:
\be
  R^{\rm SM}(D)     = 0.299 \pm 0.003, \qquad
  R^{\rm SM}(D^\ast) = 0.258 \pm 0.005.  
\en
Averaging over experimental data obtained up to
2021 gave the following results:
\be
R(D)      = 0.340 \pm 0.030, \qquad 
R(D^\ast)  = 0.295 \pm 0.014 .     
\en
It can be seen that there are deviations of about 3$\sigma$.
Taking into account reports from LHCb about new measurements performed in 2022, a new averaging was performed over all available experimental data in \cite{Iguro:2022yzr}. However, the deviation from the SM practically
did not change. The LHCb collaboration reported
measuring a similar branching ratio in semileptonic
decay $B\to J/\psi + \tau\nu_\tau$~\cite{LHCb:2017vlu}. The received data is within 2$\sigma$ deviations from the range of central values predicted
by the SM. Thus, the above results are confirmed by this independent measurement.\\

\noindent
{\bf Rare decays \boldmath{$b\to s\ell^+\ell^-$.}}

Rare decays due to neutral currents with a change in strangeness $b\to s\ell^+\ell^-$ appear only at the one-loop SM level and, therefore,
are significantly suppressed. The branchings of such
decays are at the level $10^{-6}$. However, their measurement
is available at modern accelerators, with the
exception of the currently unavailable $\tau^+\tau^-$ mode.

Branching measurements $ B\to K^{(\ast)}\ell^+\ell^-$ ($\ell= e,\mu$), as well as a number of observables arising in the angular distributions, are available. The measurement of the ratio of muon-to-electron mode branchings
is of particular interest, because in relation to
\be
R(K^{(\ast)})  = \frac{ \Br( B \to K^{(\ast)} \mu^+\mu^-)}
                    {\Br( B \to K^{(\ast)} e^+e^- )}\,,
\qquad (K^{(\ast)}=K,K^{\ast})\,,
\en
the dependence on model-dependent form factors practically disappears. These branching ratios are measured by LHCb and Belle. The experimental results obtained for several bins are consistent with the expectations of the SM at the level $2,1-2,5~\sigma$. However, for the branchings $\Br( B\to K^{\ast}\mu^+\mu^-)$, $\Br(B \to K\mu^+\mu^-)$, and $\Br(B_s \to \phi\mu^+\mu^-)$ themselves, as
well as for some observables from angular distributions,
there are discrepancies between the predictions
of the SM and measurements.

\label{sec:theories}
\section*{EFFECTIVE THEORIES}

The use of effective theories based on the construction
of effective Hamiltonian is the most convenient
way to describe weak decays of b hadrons. Their construction
within the framework of the SM is based on
several powerful methods from the arsenal of quantum
field theory. This is an operator expansion that allows
one to separate the contributions of small and large
distances; it is a renormalization group technique that
allows one to obtain numerical values of the Wilson
coefficients at energies on the order of the b-quark
mass and, at the final stage, to obtain a set of four-fermionic
operators that make it possible to describe
weak decays of hadrons. The main problem is the calculation
of the matrix elements of these operators in
the layers of the physical states of the initial and final
particles. Their calculation requires the use of nonperturbative
methods such as lattice calculations, various
sum rules, and quark models.
At the final stage, the amplitude of the weak transition
of the initial meson to the final state is written as
\bea
&&
A(f\to i)=\langle f|\,{\cal H_{\rm eff}}\,|i\rangle =
\frac{G_F}{\sqrt{2}}\lambda_{\rm CKM}\sum\limits_k
\underbrace{ C_k(\mu)}_{\rm SD} \,\,\,
\underbrace{\langle f|Q_k(\mu) |i\rangle}_{\rm LD} .
\ena
Here, $C_k(\mu)$ is Wilson coefficients and $Q_k(\mu)$ is
four-fermion operators.\\

\noindent
{\bf Covariant quark model:} 

To calculate the matrix elements of four-fermionic operators, we will use the
covariant quark model, originally formulated in \cite{Branz:2009cd}
and then developed in subsequent works. It is based on
nonlocal interpolation currents with the corresponding
quantum numbers of hadrons. Therefore, for example, the currents describing mesons, baryons,
and tetraquarks have the following form:
\bea
J_M(x) &=& \int\!\! dx_1 \!\!\int\!\! dx_2\,
F_M (x;x_1,x_2)\cdot
\bar q^a_{f_1}(x_1)\, \Gamma_M \,q^a_{f_2}(x_2)
\nn
&&\nn
J_B(x) &=& \int\!\! dx_1 \!\!\int\!\! dx_2 \!\!\int\!\! dx_3\,
F_B (x;x_1,x_2,x_3)  
\nn
&&
\times\, \Gamma_1 \, q^{a_1}_{f_1}(x_1) \, 
 \Big[ 
\varepsilon^{a_1a_2a_3}  q^{T\,a_2}_{f_2}(x_2) C \, \Gamma_2 \,  q^{a_3}_{f_3}(x_3)
              \Big]
\nn
&&\nn
J_T(x) &=& 
\int\!\! dx_1\ldots\int\!\! dx_4\, 
F_T (x;x_1,\ldots,x_4) 
\nn
&&
\times\,
\Big[ 
\varepsilon^{a_1a_2c} q_{f_1}^{T\,a_1}(x_1)\, C\Gamma_1\, q_{f_2}^{a_2}(x_2)
                  \Big]
\cdot 
\Big[
\varepsilon^{a_3a_4c} \bar q_{f_3}^{T\,a_3}(x_3)\, \Gamma_2 C\, \bar q_{f_4}^{a_4}(x_4)
\Big]
\nonumber
\ena
Vertex function $F_H (x;x_1,\ldots,x_n)$ is chosen in the
translation-invariant form
\[
F_H(x;x_1,\ldots,x_n) = \delta\left(x-\sum\limits_{i=1}^n w_i x_i\right)
\Phi_H\left(\sum\limits_{i<j} (x_i-x_j)^2\right), \quad
w_i = m_i/\sum\limits_{j=1}^n m_j,
\]
where $m_i$ is the mass of the quark described by the field $q(x_i)$. The Fourier transform of a function $\Phi_H$ is chosen
as a Gaussian exponent decreasing at infinity in
the Euclidean direction. As an example, we present
the explicit form of the matrix element that arises
when calculating the weak lepton decay of a pseudoscalar
meson:
  \[
N_c\, g_P\! \int\!\! \frac{d^4k}{ (2\pi)^4 i}\, \widetilde\Phi_P(-k^2)\,
{\rm tr} \biggl[O^{\,\mu} S_1(k+w_1 p) \gamma^5 S_2(k-w_2 p) \biggr] 
=f_P p^\mu.
\]
Here, $ S_i(k)=1/(m_i - \not k_i)$ is the Dirac quark propagator, $O^{\,\mu}$ is a weak Dirac matrix with left helicity, and $f_P$ is the calculated constant of weak lepton decay.\\

\noindent
{\bf Analysis of new physics in decays\boldmath{$B \to D^{(*)} \tau \nu_{\tau}$:}}

The possible influence of NP effects in decays $B \to D^{(*)} \tau \nu_{\tau}$ was investigated in our work \cite{Ivanov:2016qtw,Ivanov:2017hun,Ivanov:2017mrj}
within the covariant quark model. The effective Hamiltonian for the quark transition $b \to c \tau^- \bar{\nu}_{\tau}$ taking into
account the effects of NP is written in the form
\be
    {\mathcal H}_{eff} \propto G_F\,V_{cb}\,
    [(1+V_L)\mathcal{O}_{V_L}+V_R\mathcal{O}_{V_R}
      +S_L\mathcal{O}_{S_L}+S_R\mathcal{O}_{S_R} +T_L\mathcal{O}_{T_L}]
    \label{eq:Heff}
    \en
where the four-fermionic NP operators have the form
\bea
    \mathcal{O}_{V_L} &=&
    \left(\bar{c}\gamma^{\mu}P_Lb\right)
    \left(\bar{\tau}\gamma_{\mu}P_L\nu_{\tau}\right)
\qquad
\mathcal{O}_{V_R} =
\left(\bar{c}\gamma^{\mu}P_Rb\right)
\left(\bar{\tau}\gamma_{\mu}P_L\nu_{\tau}\right)
\nn
%\nn[1.5ex]
\mathcal{O}_{S_L} &=&
\left(\bar{c}P_Lb\right)\left(\bar{\tau}P_L\nu_{\tau}\right)
\qquad\qquad\,
\mathcal{O}_{S_R} =\left(\bar{c}P_Rb\right)\left(\bar{\tau}P_L\nu_{\tau}\right)
\nn
%\nn[1.5ex]
\mathcal{O}_{T_L} &=&
\left(\bar{c}\sigma^{\mu\nu}P_Lb\right)
\left(\bar{\tau}\sigma_{\mu\nu}P_L\nu_{\tau}\right)
\ena
Here $\sigma_{\mu\nu}=i\left[\gamma_{\mu},\gamma_{\nu}\right]/2$, \quad $P_{L,R}=(1\mp\gamma_5)/2$, $V_{L,R}$, $S_{L,R}$, and $T_L$ are the complex Wilson coefficients governing
the NP. In the SM we have $V_{L,R}=S_{L,R}=T_L=0$. It is assumed that all neutrinos
have left-handed helicity. It is also assumed that NP
affects only the third generation of leptons.
If we assume that, in addition to the SM contribution,
only one of the NP operators is turned on simultaneously
and the NP affects only the tau modes, we
can describe the experimental data on $R(D^{(\ast)})$ with the following values of the new Wilson coefficients:
\bea
V_L &=&-1.33+i\,1.11,\qquad V_R =0.03-i\,0.60,
\nn
S_L &=&-1.79-i\,0.22, \qquad T_L =0.38-i\,0.06.
\ena
Note that the operator $\mathcal{O}_{S_R}$ is excluded at the level of deviations $2\sigma$.

\label{sec:decays}
\section*{DECAYS $B_c\to J/\psi + \bar\ell\nu_\ell$ AND $B_c\to J/\psi + K(\pi)$}

The LHCb collaboration reported relationship measurements for the following branchings \cite{Aaij:2013vcx,Aaij:2016tcz,LHCb:2017vlu}:
\bea
\mathcal{R}_{K^{+}/ \pi^{+}} &=&
\frac { \mathcal{B}(B_c^+ \rightarrow J/\psi K^+)}
      {\mathcal{B}(B_c^+ \rightarrow J/\psi \pi^+)}
=\left\{
\begin{array}{l}
  0.069 \pm 0.019 ({\rm stat}) \pm 0.005 ({\rm syst})  
    \\
    0.079 \pm 0.007 ({\rm stat}) \pm 0.003 ({\rm syst})  
\end{array}
\right.
\nn
\mathcal{R}_{ J/\psi} &=&
\frac { \mathcal{B}(B_c^+ \rightarrow J/\psi \tau^+ \nu_{\tau})}
      {\mathcal{B}(B_c^+ \rightarrow J/\psi \mu^+ \nu_{\mu})}
      = 0.71 \pm 0.17 (\rm stat) \pm 0.18 (\rm syst)
\ena
%%%%%%%%%%%%%%%%%%%%
\begin{figure} %[H]
  \begin{center}
\begin{tabular}{lr}
\includegraphics[width=0.45\textwidth]{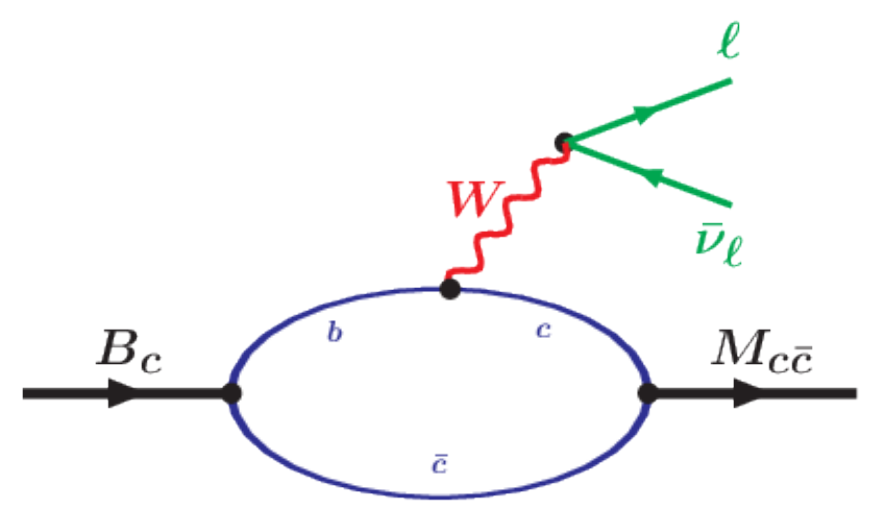} &
\includegraphics[width=0.45\textwidth]{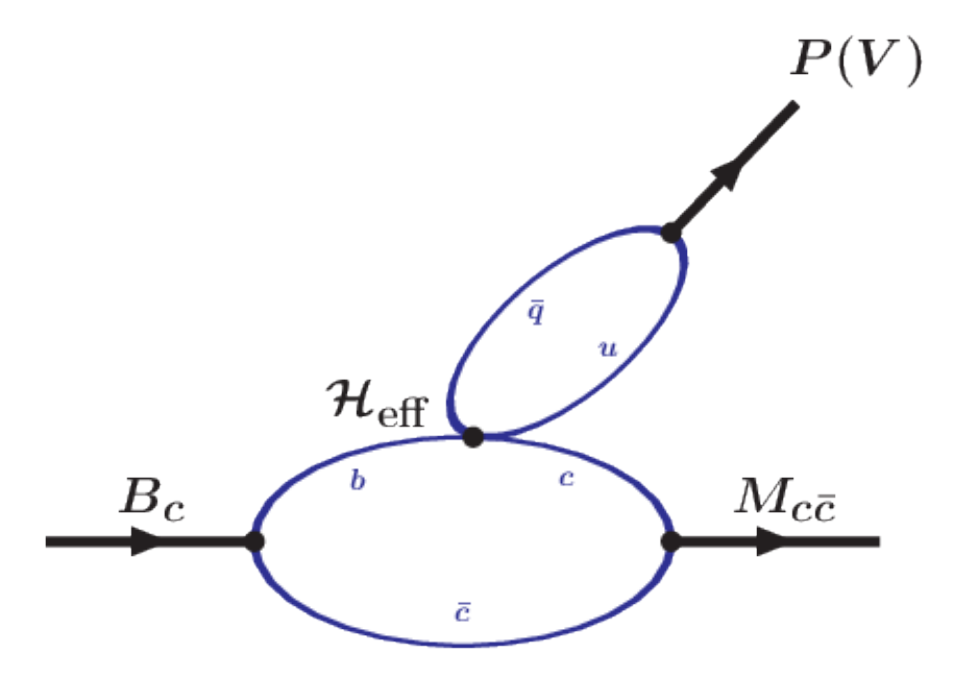} \\
\end{tabular}
\end{center}
  \caption{
Graphical representation of semileptonic and nonleptonic decays of a $B_c$ meson.}    
\end{figure}
%%%%%%%%%%%%%%%%%
\begin{figure} %[H]
\centering
\begin{tabular}{lr}
\hspace*{-1cm}  
\includegraphics[width=0.50\textwidth,height=0.27\textheight]{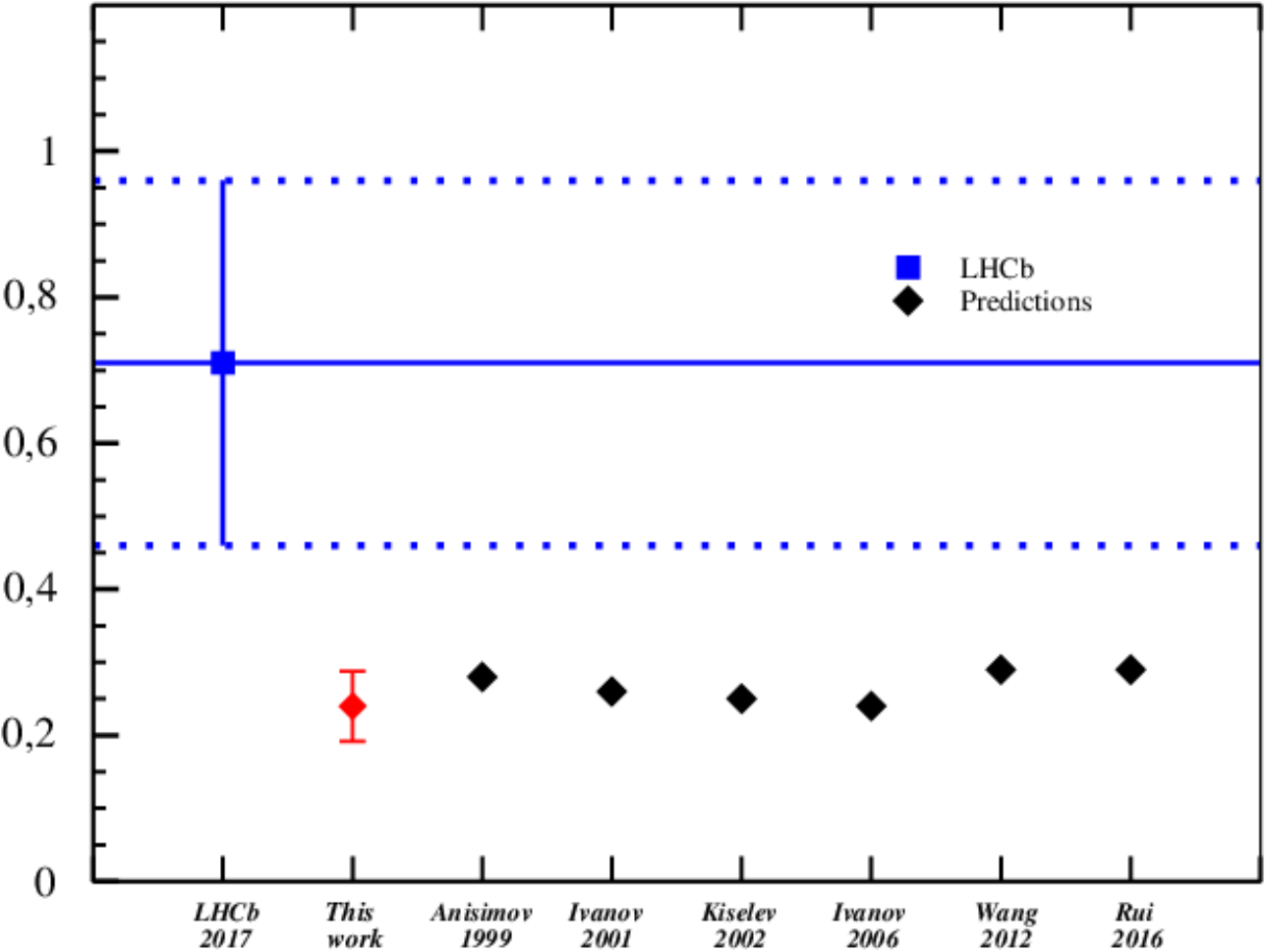} &
\includegraphics[width=0.50\textwidth,height=0.27\textheight]{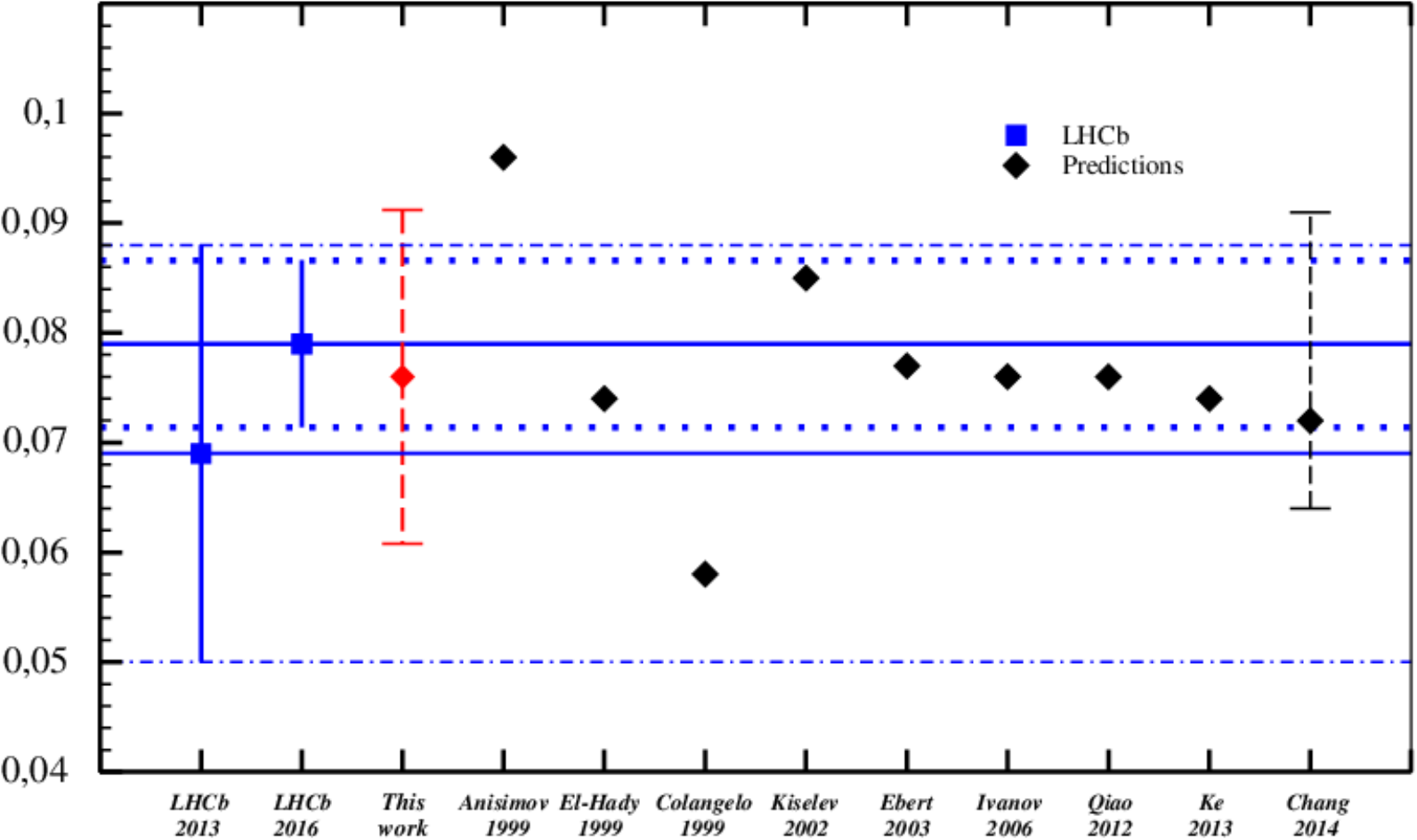} \\
\end{tabular}
  \caption{Left panel: comparison of theoretical predictions for the ratio $\mathcal{R_{J/\psi}}$ with LHCb data~\cite{Aaij:2017tyk}. The solid line is the central experimental value; the dotted lines are the experimental error bar. Right panel: comparison of theoretical predictions for the ratio $\mathcal{R_{K^{+}/\pi^{+}}}$ with LHCb data~\cite{Aaij:2013vcx,Aaij:2016tcz}. Two solid lines are central experimental values, dashed-and-dotted lines are the experimental
error bar from \cite{Aaij:2013vcx}, and dotted lines are the experimental error bar from \cite{Aaij:2016tcz}. }    
\end{figure}
%%%%%%%%%%%%%%%%%%%%%%%%%%%%%%%%%%%%%%%%%%%%%%%%%%%
In work \cite{Issadykov:2018myx}, these ratios were calculated in of the SM using the necessary form factors calculated in the covariant quark model. The results were compared
with other theoretical approaches. It turned out that
the theoretical predictions of the ratio $\mathcal{R}_{J/\psi}$ were more than 2$\sigma$ less experimental data. This may indicate the possibility of NP effects in this decay, by analogy
with the relation $\mathcal{R}_{D^{(\ast)}}$. At the same time, predictions for the ratio $\mathcal{R}_{K/\pi}$ are in good agreement with the experimental data. This may be a fairly convincing indication that the possible effects of NP manifest
themselves in the lepton sector, leading to a violation
of lepton universality, rather than in the hadron sector.\\

\noindent
    {\bf Manifestation of new physics in decays
\boldmath{$B_c \to (J/\psi,\eta_c)\tau\nu$:}}

Work~\cite{Tran:2018kuv} provided a detailed analysis of the decays $B_c \to (J/\psi,\eta_c)\tau\nu$ taking into account the NP operators. Constraints on the Wilson coefficients in the effective Hamiltonian equation~(\ref{eq:Heff}) and taking into account the NP effects in the tauon sector can be obtained with the simultaneous use of experimental data for the branching ratios $R_D=0,407\pm 0,046$ and
$R_{D^\ast }=0,304\pm 0,015$~\cite{Amhis:2016xyh} and $R_{J/\psi}=0,71\pm 0,25$~\cite{Aaij:2017tyk}. It should be noted that, in the SM, our calculation gives $R_D=0,267$, $R_{D^\ast}=0,238$, and $R_{J/\psi}=0,24$. We take into account the theoretical error in $10\%$ for our ratios. In addition, we assume the dominance of only one NP operator in addition to the SM contribution, which means that only one NP Wilson
coefficient is considered at any time.
\begin{figure} %[H]
\begin{tabular}{ccc}
\includegraphics[scale=0.3]{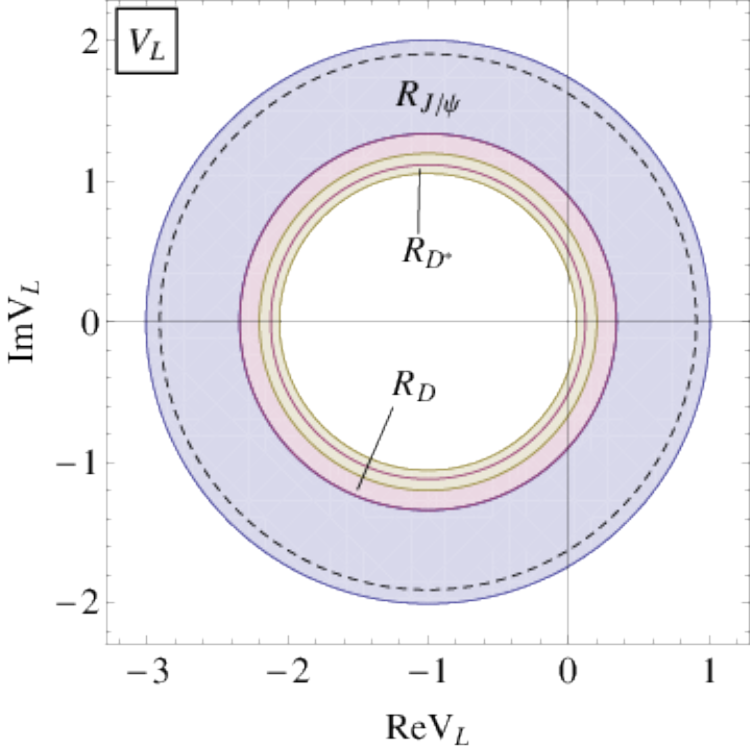}
&
\includegraphics[scale=0.3]{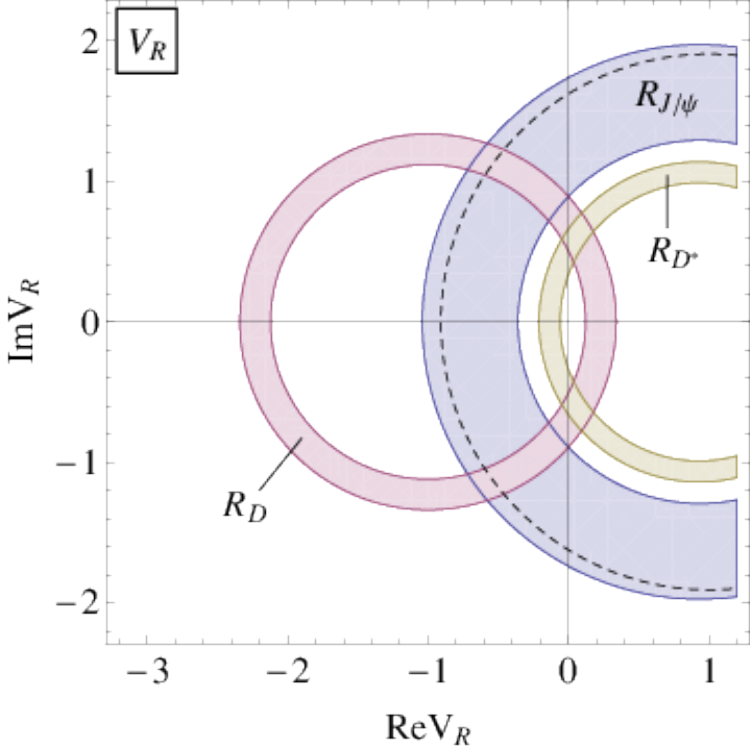}
& 
\includegraphics[scale=0.3]{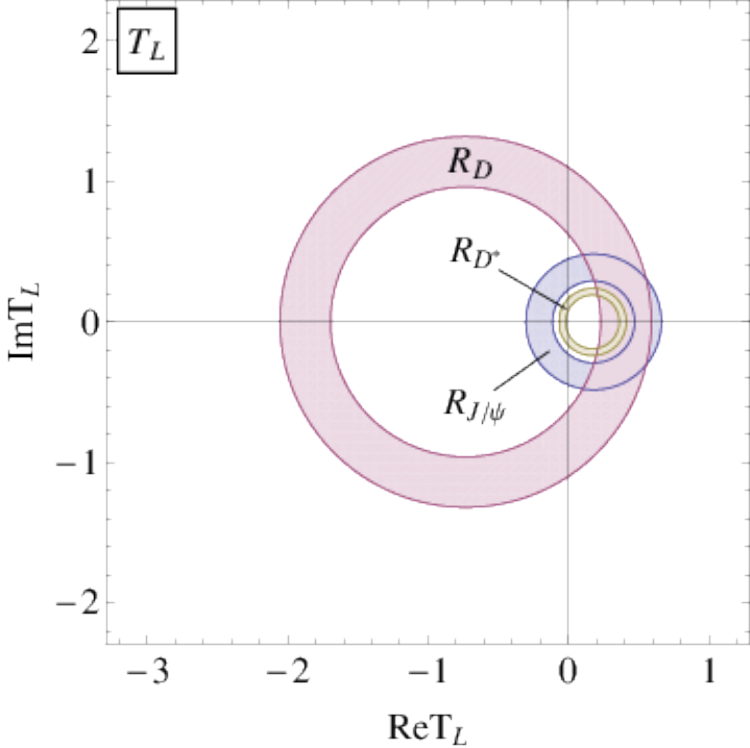}
\\
\includegraphics[scale=0.3]{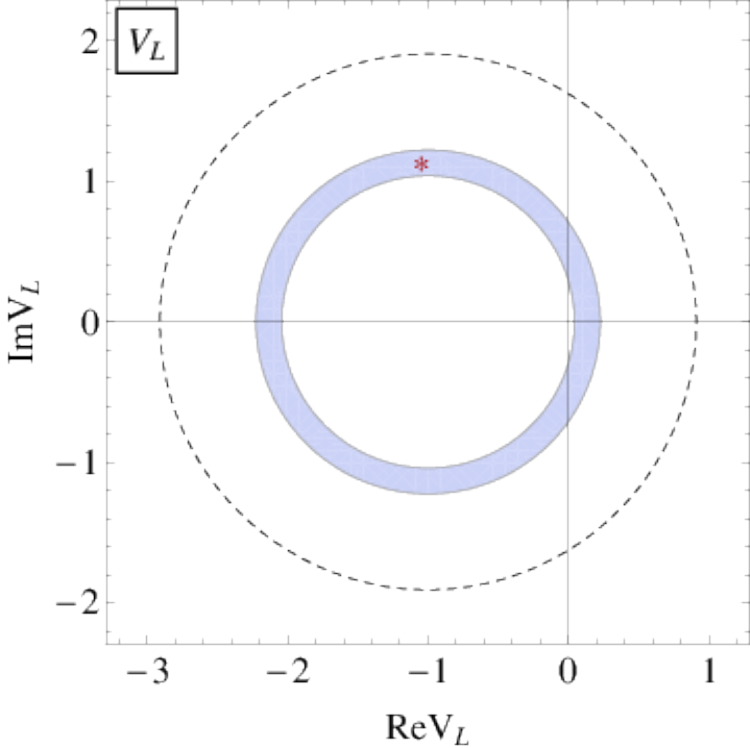}
&
\includegraphics[scale=0.3]{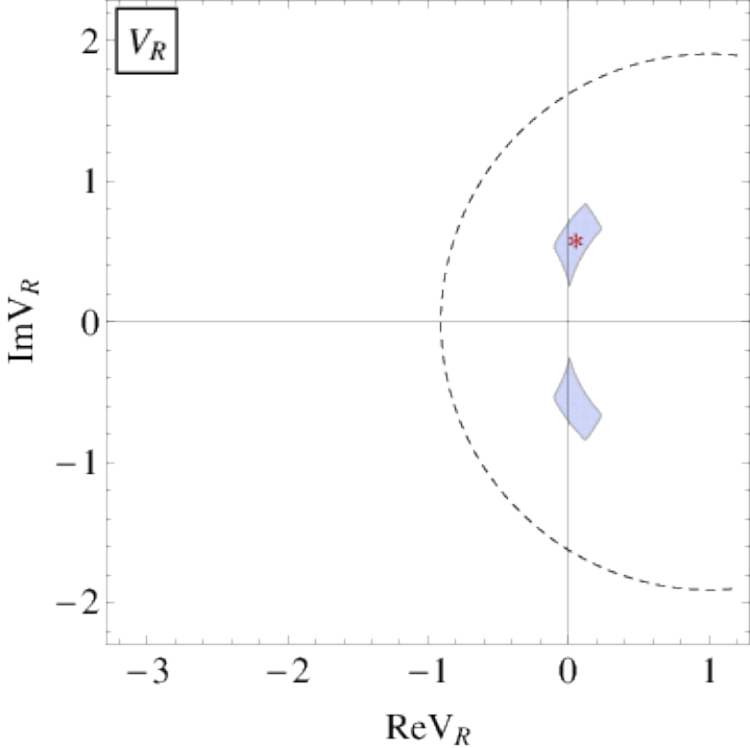}
&
\includegraphics[scale=0.3]{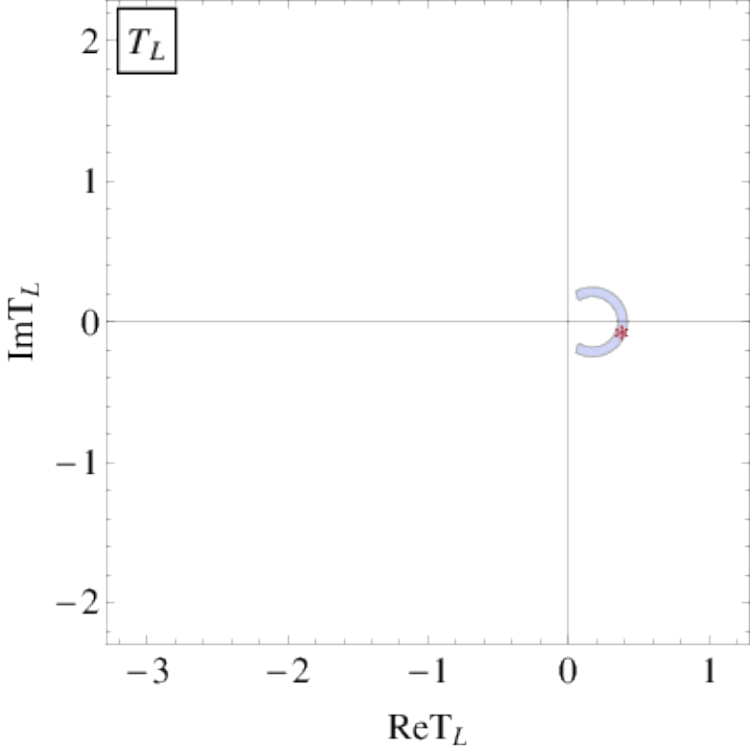}
\end{tabular}
\caption{Restrictions on Wilson coefficients $V_R$, $V_L$ and $T_L$ from ratio dimensions $R_{J/\psi}$, $R_D$ and $R_{D^\ast}$ within $1\sigma$ (top panels) and $2\sigma$ (bottom panels), as well as from the restrictions on branching $\mathcal{B}(B_c \to \tau\nu)$ (dashed curve).}
\label{fig:constraintVT}
\end{figure}

The top panels of Fig.~\ref{fig:constraintVT} present the restrictions on the vector $V_{L,R}$ and tensor $T_L$ Wilson coefficients.
Within $1\sigma$ there is no place for these coefficients.
Moreover, they are excluded mainly due to an additional restriction from $R_{J/\psi}$, not from $\mathcal{B}(B_c \to \tau\nu)$.
This is true in case $T_L$, since the operator $\mathcal{O}_{T_L}$ does not affect $\mathcal{B}(B_c \to \tau\nu)$. The bottom panels in Fig.~\ref{fig:constraintVT} show the allowed areas for $V_{L,R}$ and $T_L$ within $2\sigma$. In
each allowed area, $2\sigma$ we find the best value for each NP connection. The best fit is achieved with the following values $V_L =-1,05+i1,15$, $V_R =0,04+i0,60$, $T_L =0,38-i0,06$. In the figure, these values are
marked with an asterisk.
\begin{table} %[H] 
\begin{center}
\begin{tabular}{ccc}
\hline
&\quad  $<R_{\eta_c}>$ \qquad 
&\quad  $<R_{J/\psi}>$ \qquad   
\\
\hline
SM &\quad $0.26$\quad &\quad $0.24$\quad \\
$V_L$
&\quad $(0.28,0.39)$\quad
&\quad $(0.26,0.37)$\quad
\\
$V_R$
&\quad $(0.28,0.51)$\quad
&\quad $(0.26,0.37)$\quad
\\
$T_L$
&\quad $(0.28,0.38)$\quad
&\quad $(0.24,0.36)$\quad
\\
\hline
\end{tabular}
\caption{Ratios averaged over $q^2$ branchings in the SM and in the presence of NP.}
\label{tab:ratio}
\end{center}
\end{table}
Ratios averaged over $q^2$ branchings $R_{J/\psi}$ and $R_{\eta_c}$ are shown in Table~\ref{tab:ratio}. The row labeled SM contains our predictions obtained within the SM using the form
factors computed in our model. Predicted ranges of
ratios in the presence of NP are given according to
allowed areas $2\sigma$ NP coefficients shown in Fig.~\ref{fig:constraintVT}. It can be seen that the most noticeable effect is given by
the operator $\mathcal{O}_{V_R}$, which can increase the average ratio $<R_{\eta_c}>$ twofold.

\end{document}